# First-Principles Investigation of Mechanical, Lattice Dynamical, and Thermodynamic Properties of $BaTiO_3$ Polymorphs

Arpon Chakraborty[a], M. N. H. Liton[a,b], M. S. I. Sarker[a], M. M. Rahman[a] and M. K. R. Khan[a]*

[a]Department of Physics, University of Rajshahi, Rajshahi-6205, Bangladesh

[b]Department of Physics, Begum Rokeya University, Rangpur-5403, Bangladesh

## Abstract

$BaTiO_3$ (BTO) is one of the most interesting classes of perovskite materials. The present study has been complied to explore some physical properties such as mechanical, vibrational, thermo-physical, and temperature dependent thermodynamic properties of $BaTiO_3$ polymorphs comprehensively using first principles calculations based on density functional theory (DFT). All the polymorphs are found to be mechanically stable. The polymorphs are elastically anisotropic, machinable and have high hardness and toughness. The cubic phase possesses brittle nature while the other phases show ductile character. The high melting point of the polymorphs reveals that they can be used in tough situations. Also, three of the polymorphs can be used as thermal barrier coating. Moreover, we have also calculated the lattice dynamics and found improved results compared to the available results in the literature. In addition, the temperature and pressure dependent thermodynamic parameters of the polymorphs are evaluated and analyzed for the first time using the quasi-harmonic Debye model. The thermodynamic properties suggested that all phases would be good choices for application in the fields of automobiles, cooling systems, thermal electronic devices, thermal exchangers, and space crafts.

*Corresponding author's e-mail: mfkrkhan@yahoo.com

## 1. Introduction

The mechanical characteristics of a material are very important when attempting to characterize solids because they describe how a material reacts to being deformed. When seen from an atomistic perspective, the primary causes of elasticity are determined from the analysis of their interatomic bonding and the bonding environment. Mechanical properties indicate a compound's durability, damage resistance, and its suitable area of applications. Elastic anisotropy strongly correlated with dislocation dynamics, fracture behavior, and structural nanoscale textures. Elastic tensors are used to determine a number of thermal properties, including the Debye temperature, melting temperature, Grüneisen parameter, thermal expansion coefficient, heat capacity and thermal conductivities, of materials [1–5]. In the geophysics field, elastic properties are also widely used, since acoustic velocities may be used to analyze seismic data [6,7]. The mechanical properties of a material under the influence of external stress may be described using second-order elastic constants. The solid-solid transitions in the context of polymorphism are another significant feature of materials. Various polymorphs of the same molecular crystal have varying interaction energies, therefore other polymorphs have a tendency to transform into the most stable polymorph, which has the lowest free energy. The temperature and pressure have a great consequence on how phase transition occur in a material. A poor knowledge of the mechanochemical process at the atomistic level makes it difficult to predict and manage the phase transition [5]. The elastic constants of molecular crystals may be experimentally estimated using ultrasonic methods based on elasto-dynamics [8]. From the above discussion, it can be concluded that the calculation of mechanical properties is very crucial for its applications especially for the polymorphic structure. In this regard, the DFT-based first-principles methods have been extensively provide details knowledge about the optimal mechanical and thermodynamic properties for industrial applications. This will helpful for experimental studies with minimum cost.

By taking into account the above discussion, the present study, we focus on the deferent thermo-physical properties of Barium titanate ($BaTiO_3$) perovskite material. this material shows strong polymorphism depending on various conditions such as temperature and pressure. It has been employed different device applications, such as optical data storage, capacitors, detectors, sensors, lasers, phase conjugated mirrors, and nonlinear optical devices [9–13]. There is another interesting fact that $BaTiO_3$ has many different temperature phases. A tetragonal phase of $BaTiO_3$ (BTO) has been observed at 403–396 K by X-Ray Diffraction (XRD) and Electron Spin Resonance (ESR) methods [14]. Thermally induced magnetic transition, which is accompanied by brief trigonal crystalline shape change into a high temperature rhombohedral phase, occurs at 183 K [15]. Powder characterization revealed that cubic phase transforms into tetragonal phase with similar conditions [15]. High temperature also leads to the formation of a hexagonal polymorph associated with perovskites [16]. As a consequence, the researchers conducted studies on $BaTiO_3$ in its cubic [17,18], rhombohedral [19], orthorhombic [19,20], tetragonal [19,21], and hexagonal [16] phases. In our previous studies [22], we explored the structural, electronic and optical properties of five different phases of $BaTiO_3$. They are used in microelectronics as non-linear optical detecting devices, dielectrics, substrates for superconducting materials, and capacities for non-volatile ferroelectric memory (FERAM). Their simple crystalline structure and wide range of low-symmetry phases make them particularly interesting candidates for improving our knowledge of ferro-elelectric (FE) and anti-ferroelectroelctric phase transitions, including first and second order transitions. Besides, ceramics based on $BaTiO_3$ have numerous applications because of their high dielectric constants. It is also frequently used in the electronic industry as a multilayer ceramic capacitor (MLCC) because to their high capacitance, compact size, and dependability. Furthermore, the absence of lead in $BaTiO_3$ based ceramics makes them an attractive alternative to other piezoelectric ceramics for usage in actuators and sensors. But, the major difficulties of these applications are the thermal management because large amount of heat would be produced during their operation. Therefore, the mechanical and thermodynamic properties of the different phases of $BaTiO_3$ could play a vital role of their applications.

To our best knowledge, there are a few of the physical properties, such as structural properties, electronic band structure, density of states, phonon spectra, Debye temperature, etc. have been studied for some of the $BaTiO_3$ polymorphs so far, but not for all the $BaTiO_3$ polymorphs. Besides, a number of mechanical and thermophysical properties have not been investigated thoroughly yet for any of the phases. Earlier studies show the dynamical instability of the phonon dispersion and phonon density of states curves for all but the tetragonal phase [23,24]. Since $BaTiO_3$ is a ceramic material and many of the applications of ceramic materials are under high temperature and pressure. Therefore, it is also necessary to investigate the thermodynamic properties under elevated temperatures and pressures. Hence, there are much room to study the physical properties of those $BaTiO_3$ polymorphs, which are not explored elaborately. Here we studied the temperature and pressure dependent thermodynamic properties of the polymorphs for the first time. Furthermore, we have studied the mechanical, vibrational, thermophysical, temperature and pressure dependent thermodynamic properties of $BaTiO_3$ polymorphs. We found much improved results.

## 2. Methodology

DFT based first principles calculations of crystalline solids were carried out by means of plane-wave pseudopotential approximation using the well-known Cambridge Serial Total Energy Package (CASTEP) code [25,26]. Herein, the optimized lattice parameters of $BaTiO_3$ polymorphs are used to calculate elastic properties. Details of optimized conditions, crystal structure and lattice parameters are available in elsewhere [22]. We used the finite-strain theory [27] to calculate the elastic constants $C_{ij}$. By this theory, a given set of identical deformations (strains) is applied to the conventional unit cell and it accepts the relaxation of the atomic degrees of freedom. The resulting external stresses are then evaluated from it. For each strain $\delta_j$ applied to the unit cell, the stress tensor contains six stress components, $\sigma_{ij}$. A set of linear equations $\sigma_{ij} = C_{ij}\delta_j$ are then solve to estimate the elastic constant, $C_{ij}$. This procedure has already been applied by a number of previous studies to calculate the elastic properties [28–31]. The 2D and 3D plots of the shear modulus, Young modulus, Poisson's ratio, and linear compressibility are generated using ELATE: Elastic tensor analysis software [32] that are provided in the supplementary file.

The phonon dispersion and phonon density of states are determined using the finite displacement supercell method which is based on density functional perturbation theory (DFPT) [33–35]. The supercell dimension is fixed with the cutoff radius of 3.0 Å, which results in a supercell of volume four times that of the unit cell. For phonon calculations, norm-conserving pseudopotentials with cut-off energy 800 eV are used for all atoms. 4×4×4, 4×4×4, 8×8×8, 4×4×4, and 8×8×3 k-point mesh in the Monkhorst-Pack grid scheme [36] were used for Brillouin zone sampling for the cubic, rhombohedral, orthorhombic, tetragonal, and hexagonal phases of $BaTiO_3$.

The direct explanation of many solid-state phenomena depends critically on the thermal characteristics of crystalline materials. The properties such as bulk moduli, Debye temperatures, specific heats, volume thermal expansion coefficients, entropies, and internal energies of $BaTiO_3$ polymorphs have been evaluated for the first time at different temperatures and pressures.

The temperature dependence of the energy can be written by the work of Baroni as [33]:

$$E(T) = E_{tot} + E_{zp} + \int \frac{\hbar\omega}{\exp\left(\frac{\hbar\omega}{kT}\right) - 1} F(\omega)\, d\omega$$

where $E_{ZP}$ is defined as the zero-point vibrational energy, $k$ is Boltzmann's constant, $\hbar$ is Planck's constant, and $F(\omega)$ is the phonon density of states. $E_{ZP}$ can be evaluated as:

$$E_{ZP} = \frac{1}{2}\int F(\omega)\hbar\omega d\omega$$

$F$ is the vibrational contribution to the free energy and it is:

$$F(T) = E_{tot} + E_{ZP} + kT\int F(\omega) ln\left[1 - exp\left(-\frac{\hbar\omega}{kT}\right)\right] d\omega$$

$S$ is the vibrational contribution to the entropy and it can be written as:

$$S(T) = k\left\{\int \frac{\frac{\hbar\omega}{kT}}{\exp\left(\frac{\hbar\omega}{kT}\right) - 1} F(\omega)\, d\omega - \int F(\omega)\, ln\left[1 - exp\left(-\frac{\hbar\omega}{kT}\right)\right] d\omega\right\}$$

And the lattice contribution to the heat capacity, $C_V$, is:

$$C_V(t) = k\int \frac{\left(\frac{\hbar\omega}{kT}\right)^2 exp\left(\frac{\hbar\omega}{kT}\right)}{\left[exp\left(\frac{\hbar\omega}{kT}\right) - 1\right]^2} F(\omega)\, d\omega$$

This gives rise to the temperature-dependent Debye temperature, $\Theta_D(T)$. In the Debye model, heat capacity is given by [37]:

$$C_V^D(T) = 9Nk\left(\frac{T}{\Theta_D}\right)^3 \int_0^{\Theta DT} \frac{x^4 e^x}{(e^x - 1)^2} dx$$

where $N$ is the number of atoms per cell. Thus, the value of the Debye temperature, $\Theta_D$ at a given temperature, $T$ is attained by calculating the actual heat capacity and then inverting it to obtain $\Theta_D$. We used the quasi-harmonic Debye model [38] to evaluate the thermodynamic properties where the temperature range was selected from 0 K to 1000 K.

## 3. Results and discussion

### 3.1 Mechanical and elastic properties

Different mechanical characteristics of a material, for example, elastic constants, machinability, hardness, brittle or ductile behavior, and anisotropy are important for technological aspects. Elastic constants are crucial in determining mechanical stability, or a material's response to a macroscopic stress. The calculated values of elastic constants of $BaTiO_3$ polymorphs are given in the Table 2. The conditions of mechanical stability are vital in determining the stability of the compound. The calculated elastic constants of different phases of $BaTiO_3$ should fulfill the Born for stability are shown in Table 1 [39].

Table 1: Stability criteria for $BaTiO_3$ polymorphs

| Phase | Criteria |
|---|---|
| Cubic | $C_{11} - C_{12} > 0$; $C_{11} + 2C_{12} > 0$; $C_{44} > 0$. |
| Rhombohedral | $C_{11} > \lvert C_{12} \rvert$; $C_{13}^2 < \frac{1}{2}C_{33}(C_{11} + C_{12})$; $C_{44} > 0$;<br>$C_{14}^2 < \frac{1}{2}C_{44}(C_{11} - C_{12}) = C_{44}C_{66}$ where $C_{66} = (C_{11} - C_{12})/2$. |
| Orthorhombic | $C_{ii} > 0$ (i = 1,4,5,6); $C_{11}C_{22} > C_{12}^2$;<br>$C_{11}C_{22}C_{33} + 2C_{12}C_{13}C_{23} - C_{11}C_{23}^2 - C_{22}C_{13}^2 - C_{33}C_{12}^2 > 0$. |
| Tetragonal | $C_{11} > \lvert C_{12} \rvert$; $2C_{13}^2 < C_{33}(C_{11} + C_{12})$; $C_{44} > 0$; $C_{66} > 0$<br>where $C_{66} = (C_{11} - C_{12})/2$. |
| Hexagonal | $C_{11} > \lvert C_{12} \rvert$; $2C_{13}^2 < C_{33}(C_{11} + C_{12})$; $C_{44} > 0$; $C_{66} > 0$ |

It can be shown from the computed data that the elastic constants meet the above stability conditions. As a result, the cubic, rhombohedral, orthorhombic, tetragonal, and hexagonal phases of $BaTiO_3$ can be considered mechanically stable.

Table 2: The calculated elastic constants $C_{ij}$ (GPa), Cauchy pressure, C″ (GPa) and tetragonal shear modulus, C′ (GPa) of the cubic, rhombohedral, orthorhombic, tetragonal, and hexagonal phases of $BaTiO_3$, respectively.

| **Constant** | Cubic | Rhombohedral | Orthorhombic | Tetragonal | Hexagonal |
|---|---|---|---|---|---|
| $C_{ij}$ | $C_{11}$ = 281.93<br>$C_{12}$ = 103.00<br>$C_{44}$ = 120.27 | $C_{11}$ = 226.06<br>$C_{12}$ = 59.49<br>$C_{13}$ = 25.00<br>$C_{33}$ = 206.13<br>$C_{44}$ = 35.73 | $C_{11}$ = 258.33<br>$C_{12}$ = 61.87<br>$C_{13}$ = 85.78<br>$C_{22}$ = 211.35<br>$C_{23}$ = - 4.18<br>$C_{33}$ = 63.84<br>$C_{44}$ = 11.03<br>$C_{55}$ = 32.58<br>$C_{66}$ = 109.16 | $C_{11}$ = 295.82<br>$C_{12}$ = 106.98<br>$C_{13}$ = 100.00<br>$C_{33}$ = 166.09<br>$C_{44}$ = 18.31 | $C_{11}$ = 240.68<br>$C_{12}$ = 77.34<br>$C_{13}$ = 65.40<br>$C_{33}$ = 279.25<br>$C_{44}$ = 58.15 |
| C" | - 17.27 | 23.76 | 50.83 | 88.67 | 19.19 |
| C′ | 89.46 | 98.23 | 83.28 | 94.41 | 81.67 |

The values of diagonal elastic constants, $C_{11}$, $C_{22}$, and $C_{33}$ measure resistivity to tensile stress along the crystallographic directions *a*, *b,* and *c,* respectively. The elastic constants $C_{12}$, $C_{13}$, $C_{44}$, and $C_{66}$, on the other hand, quantify the elasticity in shape in response to shear stress. $C_{11}(=C_{22})$ is greater than $C_{33}$ for the rhombohedral, orthorhombic, and tetragonal phases, indicating that compressibility along the *c*-axis is stronger than that along the *a/b*-axis. Hence, the strength of the bonding is greater along the *a/b*-direction. The opposite is true of the hexagonal structure because $C_{11}(=C_{22})$ is less than $C_{33}$. The elastic constants $C_{44}$ and $C_{66}$ determine the resistance of the crystal in response to share deformation in the [010] and [001] directions of the compound when a tangential stress is applied to the (100) plane. $C_{44}$ is significantly lower than each of $C_{11}$ and $C_{33}$ for all five polymorphs of $BaTiO_3$ which means that a shear stress can deform the compound more easily than a one-dimensional stress applied along crystallographic directions. The shear constant, $\left(C' = \frac{C_{11}-C_{12}}{2}\right)$ determines the dynamical stability of a crystalline material. It is also a metric for crystal stiffness due to shear deformation caused by an applied shear stress along $[1\bar{1}0]$ direction in the (110) plane. For a positive value of C', the material is said to be stable; otherwise, it is dynamically unstable. From our calculations, it can be seen that all the polymorphs are stable. Voigt [40], Reuss and Angew [41] separately provided approximations to evaluate the strain in terms of a specified stress. Hill [42,43] demonstrated that Voigt and Reuss' approximations represent the upper and lower boundaries of the elastic constants, respectively. The arithmetic means of Voigt and Reuss limits is employed in Hill's approximation to represent the actual polycrystalline constants. Hence, according to Hill approximation, the mean values of bulk ($B_H$) and shear ($G_H$) moduli are:

$$B_H = \frac{B_V+B_R}{2};\ G_H = \frac{G_V+G_R}{2}$$

The Young moduli, E, and the Poisson’s ratio, ν can be determined from B and G by applying the following relations [44,45]:

$$E = \frac{9BG}{3B+G};\ \nu = \frac{3B-2G}{2(3B+G)}$$

The bulk modulus, *B* measures the resistance of a solid to volume change caused under applied stress, whereas the shear modulus, *G* measures resistance to shear deformations caused by shear stress, or plastic deformation [46]. We have found that for all phases, *G* is less than $B$ (Fig. 1(a,b)). As a result, the applied shear component should regulate mechanical failure in $BaTiO_3$. A relatively large value of $B$ and $G$ implies that all the polymorphs are hard materials and the sequence of hardness is cubic > hexagonal > tetragonal > rhombohedral > orthorhombic (Table S1).

The Young's modulus [47], $E$ is a measure of resistance to uniaxial stress, i.e., the stiffness of the material. The calculated value of $E$ (Fig. 1c) of $BaTiO_3$ polymorphs is large which corresponds to higher stiffness of the compounds. Among them, the cubic phase possesses the highest stiffness than the others. Another parameter called lattice thermal conductivity, $k_l$ is related to $E$ by the relation $k_l \sim \sqrt{E}$ [48]. Since the value of $E$ is relatively high, hence the lattice thermal conductivity of these phases is high.

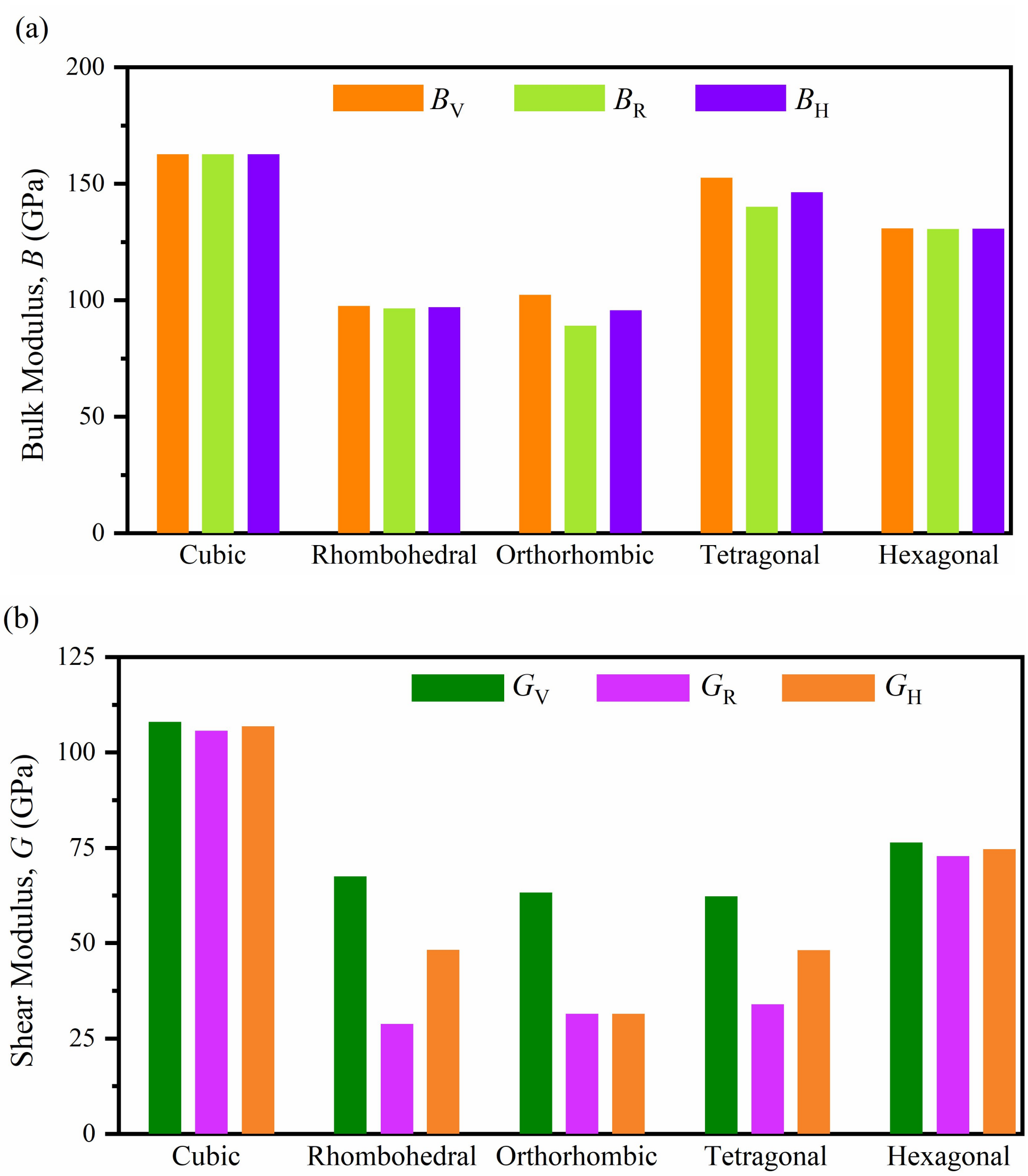

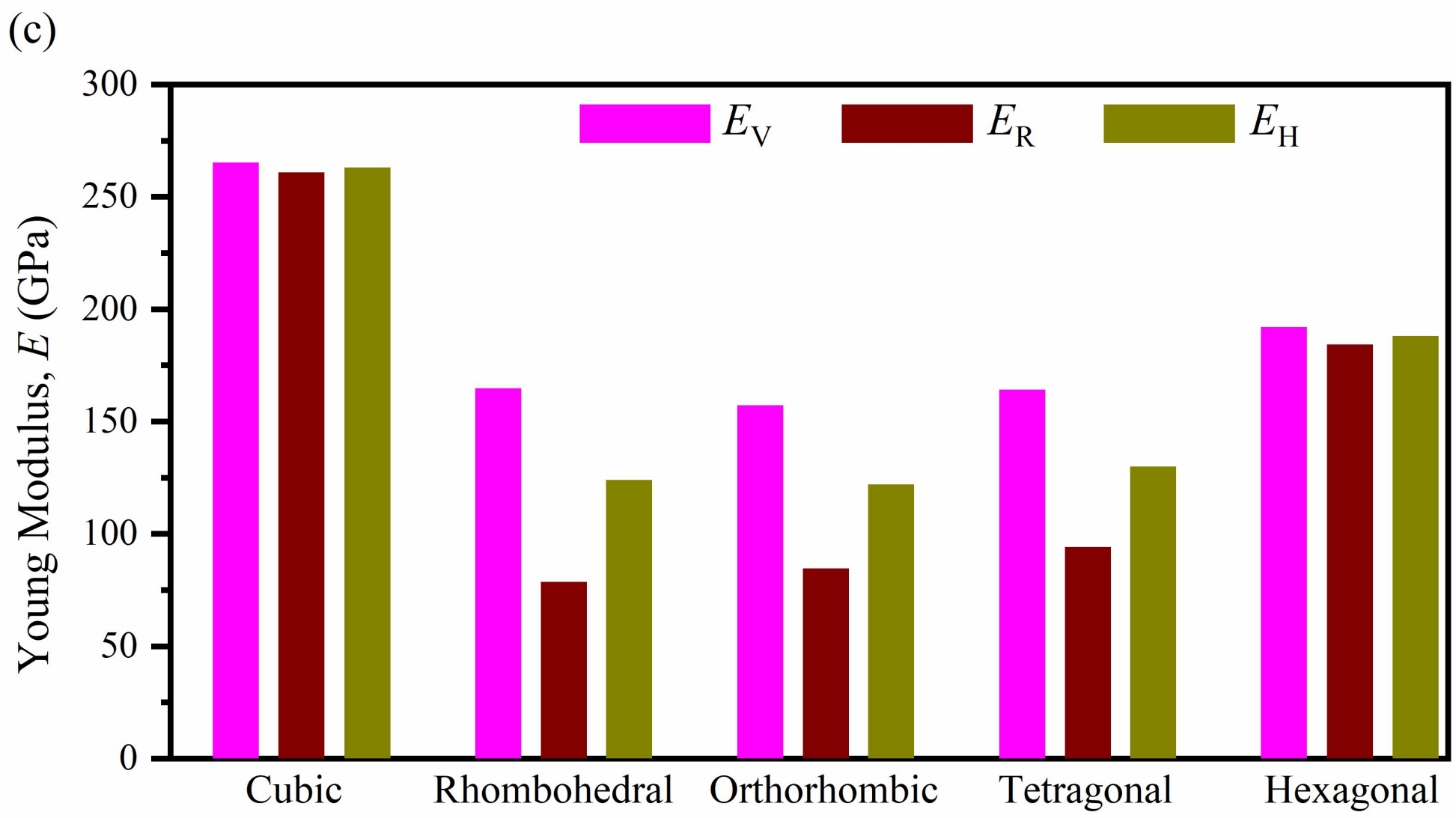

Fig. 1. Elastic moduli of different polymorphs of $BaTiO_3$

Cauchy pressure (*C''*) [49] is defined by the difference between two specific elastic constants, ($C_{12}$–$C_{44}$). Cauchy pressure can be used to determine the ductility/brittleness (failure mode) of the crystalline solids. The material is predicted to be ductile (brittle) if Cauchy pressure is positive (negative). Positive and negative Cauchy pressure also corresponds to the manifestation of the bonding (ionic/covalent) picture in a crystal system, respectively [50]. From our calculations, it can be seen that the cubic phase of $BaTiO_3$ exhibits a negative Cauchy pressure indicating brittle nature. Cauchy pressure is positive for the other four phases hence they possess ductile character. It has a positive value for metallic bond and a negative value associated with a non-central angular directed covalent bond. As a result, directional covalent bonding is expected in the cubic phase. On the other hand, the chemical bonding of the other four phases is metallic.

In practice, the majority of materials are classed as ductile or brittle. The brittleness/ductility of a compound can be estimated by the *B/G*, known as the Pugh ratio [51]. The *B/G* ratio also provides information about a material's covalent and ionic behavior based on its brittle and ductile character in solids. The limiting value of Pugh's ratio is 1.75, which separates the brittle and ductile group of materials. For brittle materials, *B/G* is less than the critical value, while for ductile materials, it is higher than the limiting value. From our present calculations, it is observed that the cubic phase possesses brittle character, and the remaining four phases i.e., rhombohedral, orthorhombic, tetragonal, and hexagonal phases demonstrate ductile nature.

The Kleinman parameter, $\xi$ which can be defined as an internal strain parameter. It explains the relative positions of the anion and cation sub-lattices throughout volume conserving distortions[52,53]. It is also a measure of a compound's resistance to stretching and bending. The Kleinman parameter ($\xi$) is estimated from the following relation [52]:

$$\xi = \frac{C_{11} + 8C_{12}}{7C_{11} + 2C_{12}}$$

The value of $\xi$ of a compound is usually in the range of zero to one ($0 \leq \xi \leq 1$), where $\xi = 0$ and $\xi = 1$ reflect the effective contribution of bond bending and bond stretching to resist externally applied stress. The computed value of $\xi$ for the cubic and tetragonal phases is ~0.5, indicating that both bond bending and stretching equally contribute to mechanical strength. The values of $\xi$ for the rhombohedral, orthorhombic, and hexagonal phases, on the other hand, are 0.41, 0.39 and 0.47, respectively, implying that bond bending contributes more to mechanical strength than bond stretching or contracting.

Another important parameter, the Poisson's ratio, $v$ is used for determining a variety of features of a compound, including its brittle/ductile nature, compressibility, and bonding force characteristics [54,55]. Poisson's ratio analogous to Pugh's ratio can be used to classify ductile and brittle materials. The threshold value to distinguish these two groups is 0.26, a value lower than this indicates brittleness of the compound and the value larger than 0.26 implies ductility of a material. Poisson's ratio provides similar results (Fig. 2) for the five polymorphs as Pugh's ratio does. From Table 3, it is seen that the calculated values of $v$ of cubic, rhombohedral, orthorhombic, tetragonal, and hexagonal phases of $BaTiO_3$ are 0.23, 0.29, 0.29, 0.35, and 0.26 respectively. The degree of directionality of the covalent bonds is also measured by Poisson's ratio. For a covalent substance, the Poisson's ratio is minimal ($v$ = 0.1), whereas, for ionic materials, it is more than or equal to 0.25 [56]. The Poisson's ratio also reflects the crystal's resistance to shear. From the obtained results, it can be seen that the covalent contribution dominates in the cubic phase, whereas the ionic contribution dominates in the other phases. It is reported that ν ranges from 0.25 to 0.50 for solids where central force interaction dominates, outside of this range non-central force become active [57]. In the cubic phase, the central force is dominant, while in the case of the rhombohedral, orthorhombic, tetragonal, and hexagonal phases, non-central force is dominant.

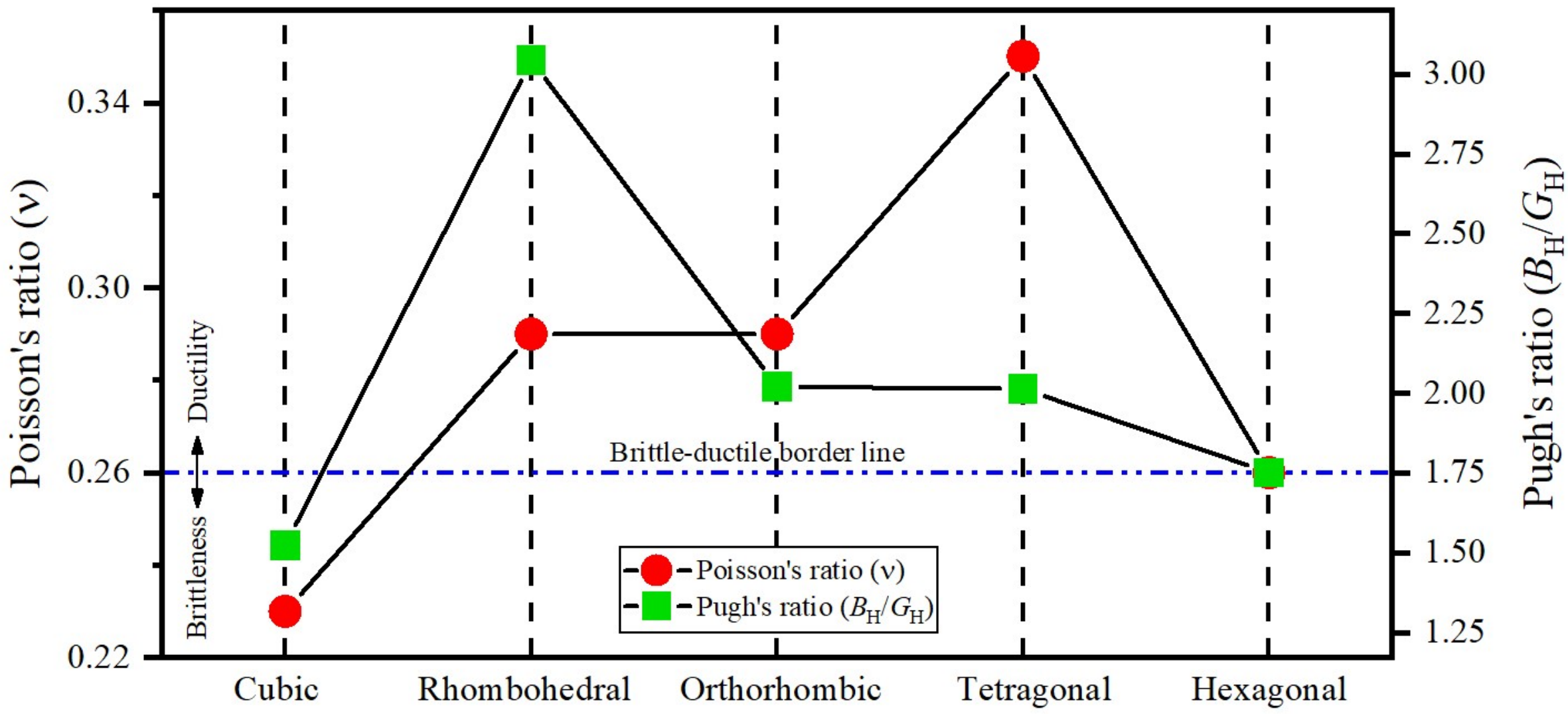

Fig. 2. Poisson's and Pugh's ratio of the $BaTiO_3$ polymorphs

The machinability index quantifies how easily a material may be machined or moulded using various cutting tools and machining methods. It is a numerical value that represents the difficulty of machining a material in comparison to a standard material under defined machining conditions. It is prejudiced by a number of factors, including the work materials' inherent properties or traits, cutting tool, the shape of the tool, the nature of tool interaction with the ask, cutting fluid, cutting circumstances, machine tool stiffness, and capacity. The machinability index, $\mu_M$ determines the cutting force, tool wear, and many other aspects of a material. The machinability index, $\mu_M$ of a material can be defined as [58]:

$$\mu_M = \frac{B}{C_{44}}$$

The plasticity [59,60] and dry lubricating properties of a substance are also measured using this parameter. A material with a high $\mu_M$ value has excellent dry lubrication qualities, low feed forces, low friction, and a high plastic strain value. From our calculation, it is observed that the machinability of $BaTiO_3$ is in the order of orthorhombic > tetragonal > rhombohedral > hexagonal > cubic phase. The high value of $\mu_M$ suggests that $BaTiO_3$ polymorphs can be cut into desired shape and hence highly applicable to designing industrial machinery.

We used an empirical approach to investigate the hardness of the polymorphs, $H_{macro}$ and $H_{micro}$ where the formulas can be expressed as [61,62]:

$$H_{macro} = 2\left[\left(\frac{G}{B}\right)^2 G\right]^{0.585} - 3$$

and

$$H_{micro} = \frac{(1-2\nu)\,E}{6(1+\nu)}$$

The values of micro and macro hardness of the present polymorphs are tabulated in Table 3. It is seen that the hardness of the cubic phase is higher than the others. $H_{micro}$ is found to be greater than $H_{macro}$ when calculated using Chen's method [61]. The differences in hardness values are caused by the parameters in the equations.

The fracture toughness, $K_{IC}$ is an essential feature in determining the mechanical characteristics of solids [63]. It gives the resistance of a solid capable of preventing the spread of an internal break. For industrial applications, a material must have high fracture toughness as well as hardness; as a result, fracture toughness prediction has piqued engineers' interest [59,64,65]. The fracture toughness, $K_{IC}$ can be determined by using the formula proposed by Niu et al. [59]:

$$K_{IC} = V_0^{\frac{1}{6}} G (B/G)^{0.5}$$

The obtained values of $K_{IC}$ for the polymorphs are listed in Table 3. From Table 3, we can again see that the value of toughness of the compound is large, and cubic and hexagonal phase's shows the highest toughness. Hence, these polymorphs can be used to design ultrahigh temperature ceramic materials against crack propagation.

Table 3: Pugh's ratio *B/G*, Poisson's ratio $\nu$, machinability index, $\mu_M$, Kleinman parameter, $\xi$, and hardness values, $H_{macro}$ (GPa) and $H_{micro}$ (GPa), and fracture toughness, $K_{IC}$ ($MPam^{0.5}$) of polymorphs of $BaTiO_3$.

| Phase | B/G | $\nu$ | $\mu_M$ | $\xi$ | $H_{macro}$ | $H_{micro}$ | $K_{IC}$ |
|---|---|---|---|---|---|---|---|
| Cubic | 1.52 | 0.231 | 1.35 | 0.50 | 15.80 | 19.19 | 2.02 |
| Rhombohedral | 3.04 | 0.287 | 2.71 | 0.41 | 5.52 | 6.85 | 1.05 |
| Orthorhombic | 2.02 | 0.288 | 8.67 | 0.39 | 5.38 | 6.70 | 1.04 |
| Tetragonal | 2.01 | 0.352 | 7.99 | 0.50 | 2.24 | 4.75 | 1.29 |
| Hexagonal | 1.75 | 0.260 | 2.25 | 0.47 | 9.95 | 11.94 | 1.52 |

### 3.2 Elastic anisotropy

It is critical to investigate the anisotropic elastic characteristics of materials because the nature and extent of anisotropy in materials are closely related to several important physical procedures such as the creation of micro-scale cracking in solids, crack motion, structural instability, phase transformations, mechanical yield points, and the growth of plastic deformations in crystals, among others. The anisotropy factors $A_1$, $A_2$ and $A_3$ also known as share anisotropy factor associated with the {100}, {010} and {001} shear planes between <011> and <010>, <101> and <001>, and <110> and <100> directions, respectively. The variation of the values of $A_i$ ($i$=1,2,3) different from unity implies the anisotropic nature of the material under consideration [66,67]:

$$A_1 = \frac{4C_{44}}{C_{11} + C_{33} - 2C_{13}}$$

$$A_2 = \frac{4C_{55}}{C_{22} + C_{33} - 2C_{23}}$$

$$A_3 = \frac{4C_{66}}{C_{11} + C_{22} - 2C_{12}}$$

Shear anisotropic factors for the five $BaTiO_3$ phases are calculated and graphically represent in Fig. 3 and also shown in the Table S3 and the results reveals that all the polymorphs are highly anisotropic.

The universal log-Euclidean index is defined as [68,69]:

$$A^L = \sqrt{\left[ln\left(\frac{B^V}{B^R}\right)\right]^2 + 5\left[ln\left(\frac{C_{44}^V}{C_{44}^R}\right)\right]^2}$$

where the Voigt and Reuss values of $C_{44}$ can be obtained from the following equations[68],

$$C_{44}^R = \frac{5}{3}\frac{C_{44}}{3(C_{11} - C_{12}) + 4C_{44}}$$

and

$$C_{44}^{V} = C_{44}^{R} + \frac{3}{5}\frac{(C_{11} - C_{12} - 2C_{44})^2}{3(C_{11} - C_{12}) + 4C_{44}}$$

We estimated $A^L$ utilizing the difference between $C^V$ and $C^R$ averaged stiffnesses, which is more appropriate. For a flawlessly isotropic crystal, $A^L$ is zero. The value of $A^L$ for the five polymorphs of $BaTiO_3$ are shown in Fig. 3 and Table S3. The values of $A^L$ varies from 0.14 to 4.03. From Table 4, it indicates that the orthorhombic and tetragonal phases have a high level of anisotropy whereas the cubic, rhombohedral, and hexagonal phases are not that much anisotropic.

For $BaTiO_3$ polymorphs, the universal anisotropy index $A^U$, equivalent Zener anisotropy measure $A^{eq}$, shear anisotropy $A^G$ (or $A^C$) and anisotropy incompressibility $A^B$ are assessed by the following relations [68,70–72]:

$$A^U = 5\frac{G_V}{G_R} + \frac{B_V}{B_R} - 6 \geq 0$$

$$A^{eq} = \left(1 + \frac{5}{12}A^U\right) + \sqrt{\left(1 + \frac{5}{12}A^U\right)^2 - 1}$$

$$A^G = \frac{G_V - G_R}{2G_H}$$

$$A^B = \frac{B_V - B_R}{B_V + B_R}$$

Ranganathan and Ostoja-Starzewski [70] introduced the universal anisotropy index, $A^U$ which gives a determine of anisotropy regardless of crystal symmetry. There are only two possible values for the universal anisotropy factor: zero or positive. Whereas $A^U$= 0 specifies isotropic crystal, and any variation from this value shows the presence and degree of anisotropy. $A^U$ deviates from zero (Fig. 3) for all the phases of $BaTiO_3$. As a result, we may state that the polymorphs under investigation have mechanical/elastic anisotropy.

For a crystal to be isotropic, $A^G$and $A^B$ must be equal to zero whereas the value other than zero measures the degree of anisotropy.

And the following equation is used to estimate the linear compressibility coefficients of the crystal [31]:

$$\frac{B_c}{B_a} = \frac{C_{11} + C_{12} - 2C_{13}}{C_{33} - C_{13}}$$

where $B_a$ and $B_c$ are the bulk moduli in the $a$-and $c$-direction, respectively. The estimated values of anisotropy factors are tabulated in Table S3. From the different anisotropy factors calculations, it can be concluded that all the $BaTiO_3$ polymorphs are elastically anisotropic. Among them, the tetragonal phase of $BaTiO_3$ possesses maximum anisotropy while the cubic one exhibits minimum anisotropy.

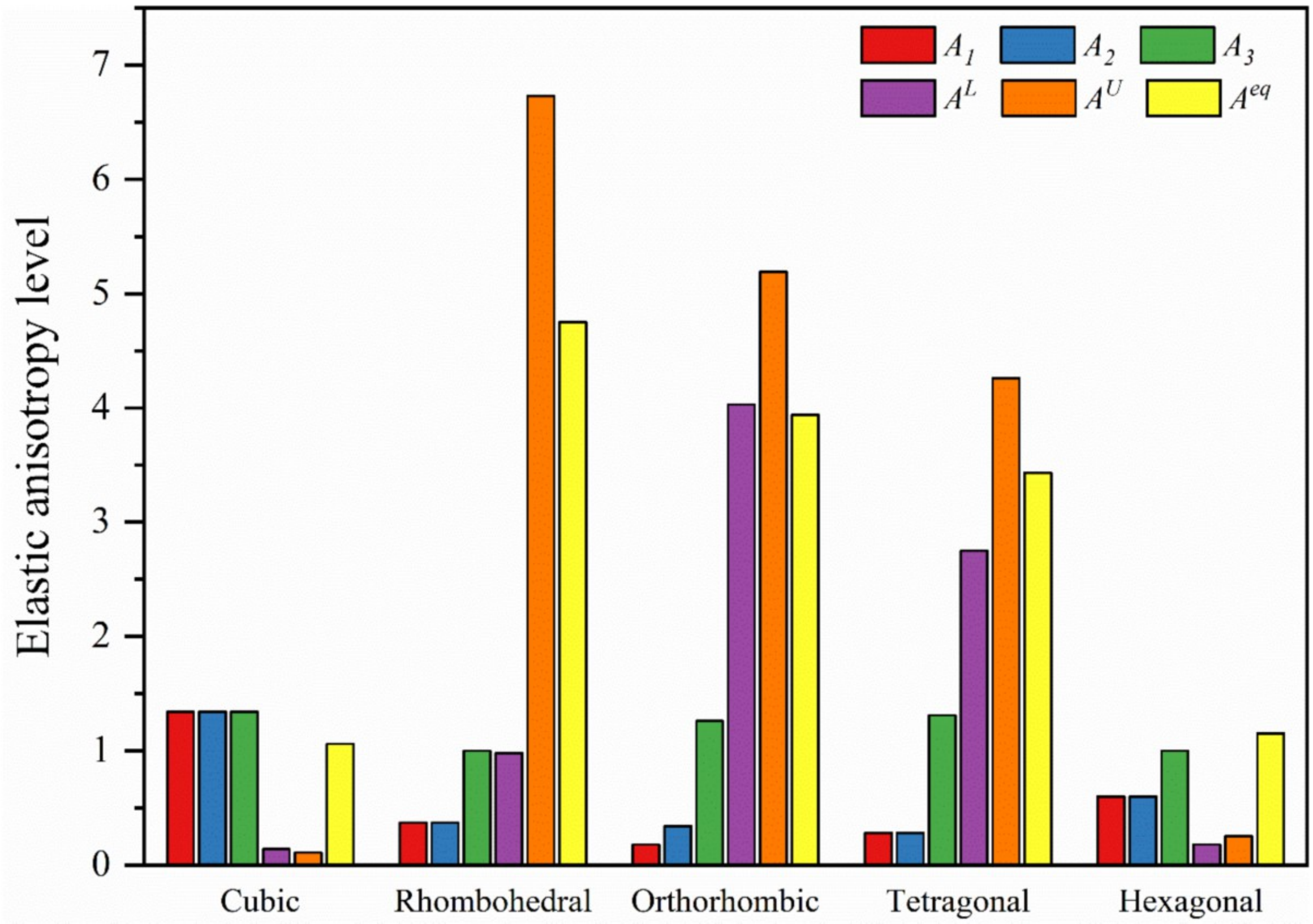

Fig. 3. Elastic anisotropy levels in different polymorphs of $BaTiO_3$

### 3.3 Acoustic velocities

The thermal and electrical conductivity of a material is influenced by another essential characteristic, sound velocity. Furthermore, in recent years, interest in understanding a material's acoustic behavior has grown significantly in modern science to design sophisticated instruments. In a crystalline material, the transverse and longitudinal sound velocities are determined from [73]:

$$v_t = \sqrt{\frac{G}{\rho}} \qquad \text{and} \qquad v_l = \sqrt{\frac{3B+4G}{3\rho}}$$

where $\rho$ is the density of the material. The calculated longitudinal velocity is ~1.5 to ~2 times higher than that of the transverse velocity, as can be seen from Table 4.

The mean sound velocity, $v_m$ can be obtained in terms of the transverse, $v_t$ and longitudinal , $v_l$ sound velocities [73]:

$$v_m = \left[\frac{1}{3}\left(\frac{2}{v_t^3} + \frac{1}{v_l^3}\right)\right]^{-\frac{1}{3}}$$

The study of a material with the unique or variable acoustic impedance as the surrounding medium has piqued interest in transducer/microphone design, noise reduction in aviation engines and the body of bullet trains, industrial plants, and a variety of undersea acoustic applications. The amount of sound (transmitted and reflected) energy at the interface is determined by the difference in acoustic impedances when sound passes through the material and then contacts another. If the two impedances are nearly identical, most of the sound is communicated; but, if the impedance difference is large, most of it is reflected back rather than sent, resulting in transmission signal loss and echo creation. The acoustic impedance, *Z*, of a material is found from [74]:

$$Z = \sqrt{\rho G}$$

According to the above equation, a material with higher density and shear modulus has a higher acoustic impedance. The calculated values of acoustic impedance for cubic, rhombohedral, orthorhombic, tetragonal, and hexagonal phases of $BaTiO_3$ are listed in Table 4.

The intensity of sound radiation is another essential parameter in acoustics. This parameter is required for creating soundboards such as the front and back plates of a guitar/violin, a harpsichord's soundboard, and a loudspeaker's panel. For a particular driving source, the radiation intensity, *I* depend on the surface velocity. This is related to the modulus of rigidity and density as [74]:

$$I \approx \sqrt{G/\rho^3}$$

where $\sqrt{G/\rho^3}$ is called the radiation factor. Instrument designers pick materials for appropriately constructed soundboards based on a high value of radiation factor. The estimated value of radiation factors for cubic, rhombohedral, orthorhombic, tetragonal, and hexagonal phases of $BaTiO_3$ are listed in Table 4. The above information will be helpful to predict the suitable area of acoustic device applications of $BaTiO_3$ polymorphs.

Table 4: Density ρ ($g/cm^3$), transverse sound velocity $v_t$($ms^{-1}$), longitudinal sound velocity $v_l$ ($ms^{-1}$), average sound wave velocity $v_m$ ($ms^{-1}$), acoustic impedance Z (Rayl=$kgm^{-2}s^{-1}$) and radiation factor $\sqrt{G/\rho^3}$ ($m^4$/kg.s) of the cubic, rhombohedral, orthorhombic, tetragonal, and hexagonal phases of $BaTiO_3$:

| **Phase** | **ρ** | $v_t$ | $v_l$ | $v_m$ | **Z (×10⁶)** | $\sqrt{G/\rho^3}$ |
|---|---|---|---|---|---|---|
| Cubic | 5.94 | 4240.86 | 7166.68 | 4697.64 | 25.19 | 0.71 |
| Rhombohedral | 5.77 | 2889.40 | 5284.80 | 3222.05 | 16.67 | 0.50 |
| Orthorhombic | 5.77 | 2864.69 | 5246.47 | 3194.82 | 16.52 | 0.50 |
| Tetragonal | 5.79 | 2881.10 | 6026.61 | 3240.08 | 16.58 | 0.50 |
| Hexagonal | 5.59 | 3621.47 | 6359.69 | 4025.30 | 20.61 | 0.64 |

Besides, we have also calculated directional sound velocity which is provided in the supplementary file.

### 3.4 Vibrational properties

Since materials are frequently subjected to time-varying mechanical stress, therefore, a dynamical stability check of the material is necessary. Besides, thermodynamic properties, such as heat capacities, thermal expansion coefficient, etc. can also be obtained from it [75–77]. The PDC along different high symmetry points in the 1st Brillouin zone has been calculated and displayed in Figs. 4(a-e). The PDCs of $BaTiO_3$ polymorphs contain three acoustic and twelve optical modes. The low frequency phonon modes are dominated by Ba atoms while Ti and O are contributing at high frequency modes. This happened because of the large atomic mass of Ba as compared to Ti and O atoms. From Figs. 4(a-e), it is seen that all the polymorphs of $BaTiO_3$ possess dynamical instability since they have negative phonon frequencies, which corresponds dynamical failure of these polymorphs due to time dependent perturbation [78].

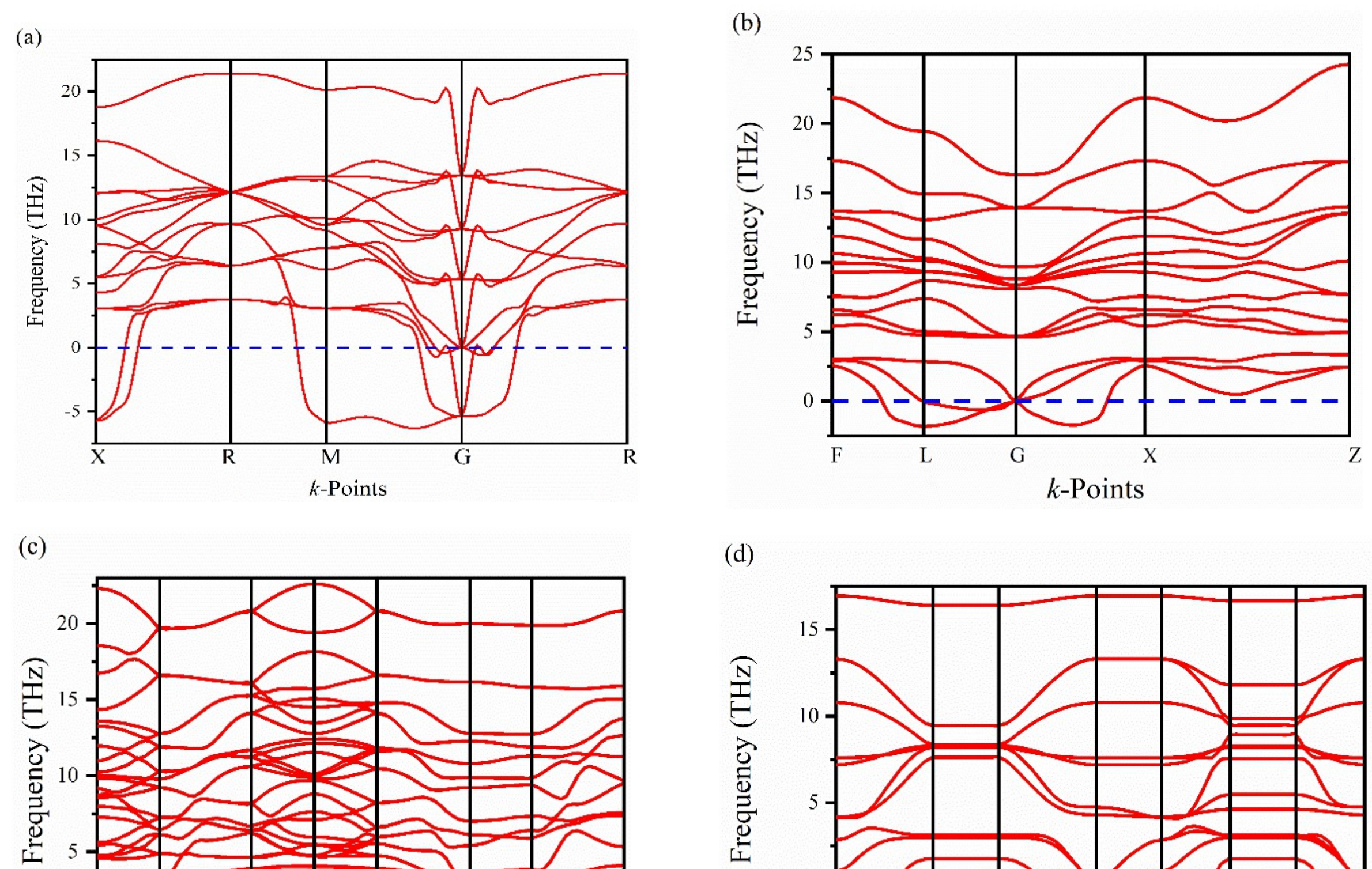

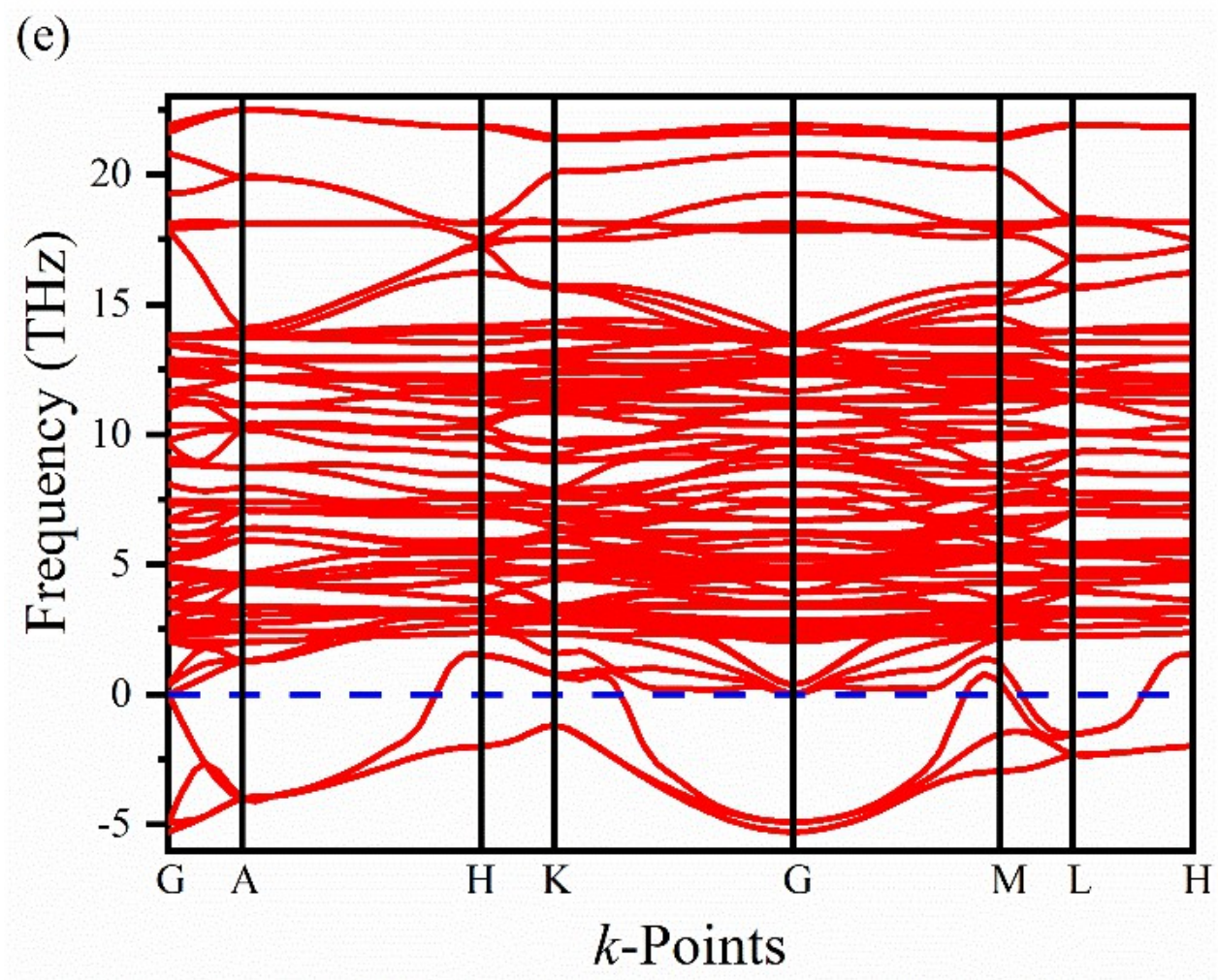

Fig. 4. Phonon dispersion curves of the (a) cubic, (b) rhombohedral, (c) orthorhombic, (d) tetragonal, and (e) hexagonal phases of $BaTiO_3$.

Zhang *et al.* [79] remarked that the Ti-O bond distortion is mainly responsible for imaginary phonon frequency. Uludoğan *et al.* [14] experimentally prove that all the polymorphs are dynamically stable meanwhile theoretical calculations demonstrate dynamical instability that might be responsible for ferroelectric phase transitions. Our results agree well with the previously available studies in the literature. Furthermore, though these imaginary frequencies indicate that the polymorphs are supposed to be not stable dynamically, but these phases are actually stable as we can see from the mechanical stability conditions (section 3.1).

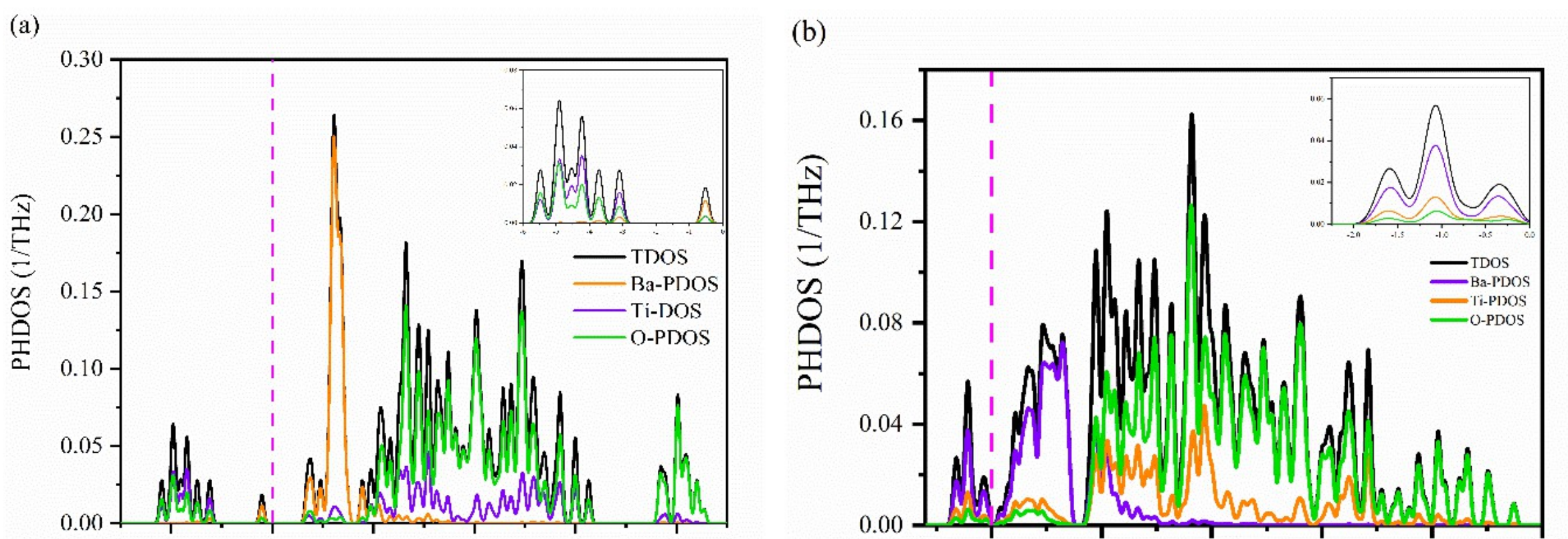

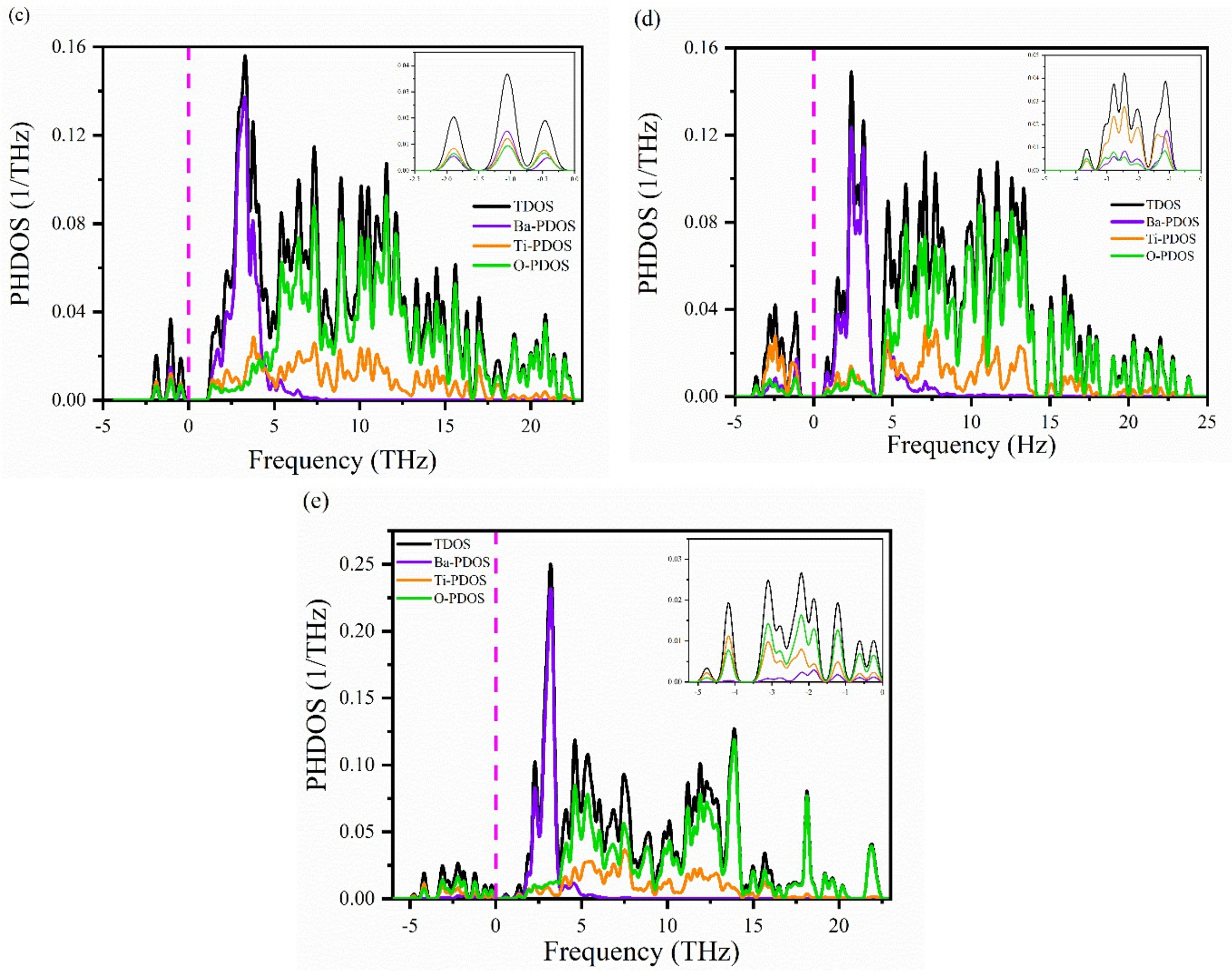

Fig. 5. Phonon density of states (PHDOS) of the (a) cubic, (b) rhombohedral, (c) orthorhombic, (d) tetragonal, and (e) hexagonal phases of $BaTiO_3$.

Figs. 5(a-e) represent the phonon total and partial density of states (PHDOS) for cubic, rhombohedral, orthorhombic, tetragonal, and hexagonal phases of $BaTiO_3$, respectively. From Figs. 5(a-e), it is depicted all the polymorphs exhibit both real and imaginary frequencies. It is also observed that all the atoms contribute to the imaginary frequencies of the vibrations and the value of negative frequency is higher for the hexagonal phase than the other phases. The degree of imaginary frequencies mainly depends on the degree of vibrations of Ti and O, and Ti-O distortions. It is eminent that most of the low symmetry perovskite phases are derived from high symmetry cubic phases by changing the orientation of octahedral units around the symmetry axes [80]. From the aforementioned discussion, we can conclude that the distortion of Ti-O is mainly responsible for negative phonon frequency.

## 3.5 Thermophysical properties

### 3.5.1 Debye temperature

The Debye temperature $\Theta_D$ is the temperature of a crystal's highest normal mode of vibration. It is a characteristic temperature of solids that allows for the calculation of several physical parameters of crystalline solids, including lattice vibration, thermal conductivity, thermal expansion, interatomic bonding, specific heat, lattice enthalpy, and melting temperature. Moreover, the energy essential to create a vacancy in metals is largely determined by their Debye temperatures. It also distinguishes between lattice dynamics and heat capacity in high- and low-temperature sections. Higher Debye temperatures are seen in materials with better interatomic bonding strength, superior hardness, high melting temperature, high mechanical wave velocity, and small average atomic mass. When the temperature is higher than $\Theta_D$, all vibration modes have an energy ~$k_BT$. High frequency modes, on the other hand, are likely to remain frozen when T < $\Theta_D$ [81], revealing the quantum nature of vibrational modes. At low temperatures, the $\Theta_D$ is estimated using the elastic constants because vibrational excitations are solely caused by acoustic vibrations at low temperatures, and the Debye temperature can be calculated using the following equations [73,82]:

$$\Theta_D = \frac{h}{k_B}\left[\left(\frac{3n}{4\pi}\right)\frac{N_\rho}{M}\right]^{1/3} v_m$$

where $h$ stands for Planck constant, $k_B$ is the Boltzmann's constant, $N_A$ is the Avogadro's number, $\rho$ indicates mass density, $M$ is the molecular weight, and $n$ is the number of atoms within the unit cell. The estimated large value of the Debye temperature (Table 5 and Fig. 6) for cubic, rhombohedral, orthorhombic, tetragonal, and hexagonal phases of $BaTiO_3$ is asserted that a higher lattice thermal conductivity is expected for these polymorphs.

### 3.5.2 Melting temperature

The melting points are used to define and evaluate the purity of organic and inorganic substances. The melting temperature ($T_m$) of a crystalline solid is related to its thermal expansion and bonding energy. The elastic constants can be used to calculate the melting temperature of solids, from the following equation [83]:

$$T_m = 354K + (4.5K/GPa)\left(\frac{2C_{11} + C_{33}}{3}\right) \pm 300K$$

A high melting point is caused by either a high heat of fusion or a low entropy of fusion or a combination of the two. The crystal phase in highly symmetrical molecules is densely packed with many active intermolecular interactions, resulting in a greater enthalpy change on melting. The calculated melting temperature of the five polymorphs of $BaTiO_3$ is listed in Table 5 and presented in Fig. 6. Due to high melting points, all the polymorphs would be potential materials for use in thermally hostile situations and are a suitable option for high-temperature applications. In addition, the higher value of melting points corresponds to the expected low thermal conductivity of these polymorphs.

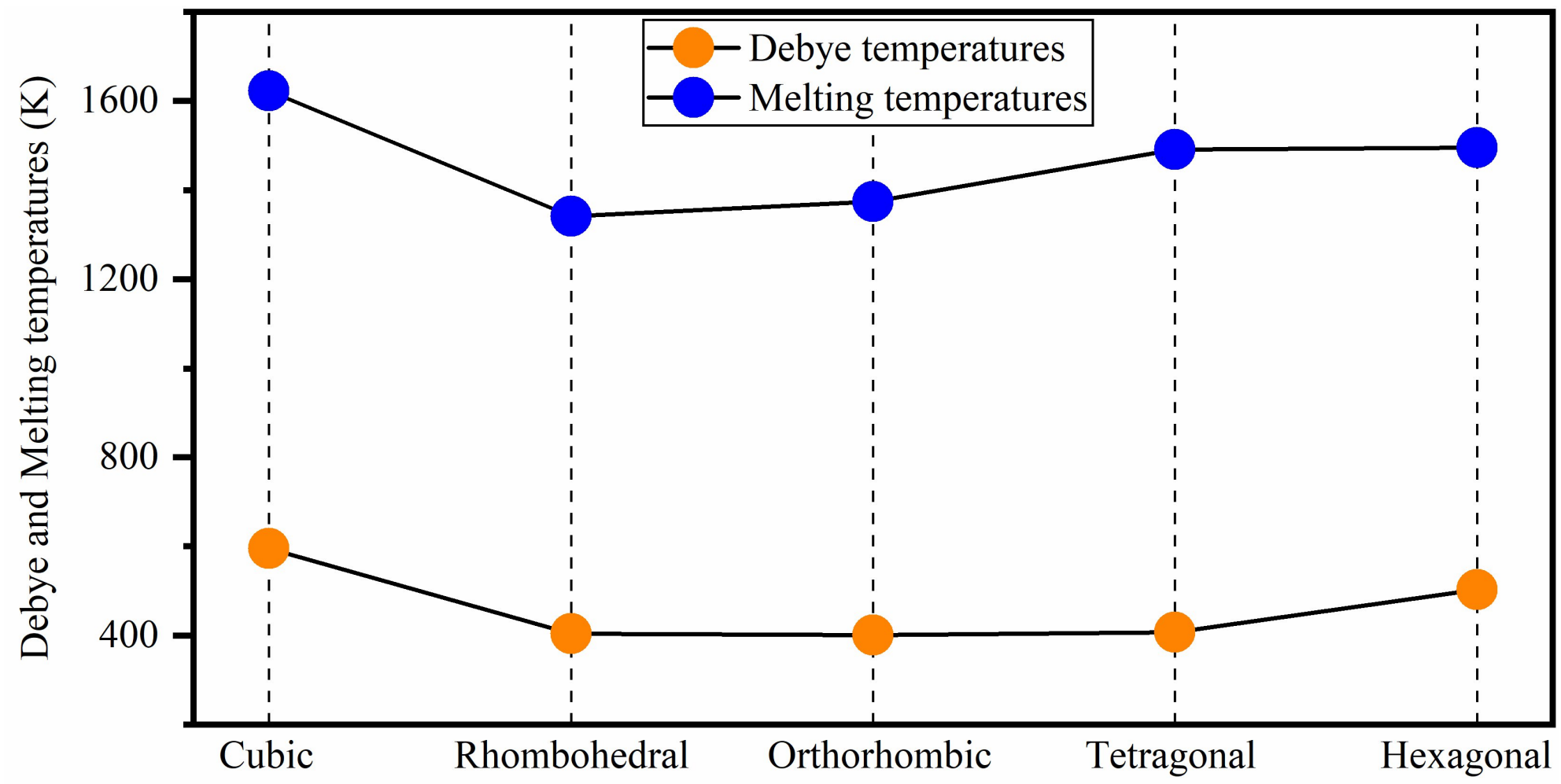

Fig. 6. Debye and melting temperature of the five polymorphs of $BaTiO_3$

### 3.5.3 Thermal expansion coefficient and heat capacity

The thermal expansion coefficient α is associated with various physical properties, including thermal conductivity, specific heat, the effective mass of the electron, and the temperature change of the energy band gap. It is a key indicator of a material's ability to act as a thermal barrier coating (TBC). It is also necessary for epitaxial development of the crystal and the minimization of adverse effects in electrical and spintronic devices. The thermal expansion coefficient of a material can be calculated as follows [74]:

$$\alpha = \frac{1.6 \times 10^{-3}}{G}$$

where $G$ is the shear modulus (in GPa). The thermal expansion coefficient of a substance is inversely proportional to its melting temperature as $\alpha = 0.02/T_m$ [74,84]. The thermal expansion coefficients for cubic, rhombohedral, orthorhombic, tetragonal, and hexagonal phases of $BaTiO_3$ are tabulated in Table 5. The higher value of melting temperature results in lower thermal expansion of the polymorphs.

Another intrinsic thermodynamic property of a molecule is its heat capacity. A higher heat capacity results in lower thermal diffusivity and higher thermal conductivity. Heat capacity per unit volume is the quantity of energy that must be given to one unit of volume of a material in the form of heat in order to create a one unit (Kelvin) increase in temperature. The heat capacity per unit volume was calculated by using the following equation [74]:

$$\rho C_P = \frac{3k_B}{\Omega}$$

Where $N = 1/\Omega$ is the number of atoms per unit volume. The heat capacity per unit volume for the five phases of $BaTiO_3$ is listed in Table 5.

Phonons, also known as quantum of lattice vibrations, play a vital part in a number of physical characteristics such as thermal conductivity, electrical conductivity, thermopower, and heat capacity. Dominant phonon wavelength, $\lambda_{dom}$which is termed as the wavelength at which phonon contributes the most to the heat transport to identify the main physical phenomenon controlling heat transport. The dominant phonon wavelength for the five polymorphs of $BaTiO_3$ at 300K has been estimated by using the following relationship [84]:

$$\lambda_{dom} = \frac{112.566 v_m}{T} \times 10^{-12}$$

where $v_m$ denotes the average sound velocity in $ms^{-1}$ and T denotes the temperature in K. The dominating phonon wavelength is longer in materials with higher average sound velocity, higher shear modulus, and lower density. The calculated value of $\lambda_{dom}$ in meter is listed in Table 5.

Table 5: The Debye temperature $\Theta_D$ (K), thermal expansion coefficient α ($K^{-1}$), the wavelength of the dominant phonon at 300K, $\lambda_{dom}$ (m), melting temperature $T_m$ (K) and heat capacity per

unit volume $\rho C_P$ (J/m$^3$.K) of the cubic, rhombohedral, orthorhombic, tetragonal, and hexagonal phases of $BaTiO_3$:

| Phase | $\Theta_D$ | α (×10$^{-5}$) | $\lambda_{dom}$ (×10$^{-12}$) | $T_m$ | $\rho C_P$ (×10$^6$) |
|---|---|---|---|---|---|
| Cubic | 594.57 | 1.50 | 196.77 | 1622.69 | 3.18 |
| Rhombohedral | 403.78 | 3.32 | 134.96 | 1341.38 | 3.08 |
| Orthorhombic | 400.37 | 3.38 | 133.82 | 1374.75 | 3.08 |
| Tetragonal | 406.61 | 3.33 | 135.72 | 1490.58 | 3.10 |
| Hexagonal | 502.17 | 2.14 | 168.61 | 1494.92 | 3.04 |

### 3.5.4 Minimum thermal conductivity

At higher temperatures, the minimum thermal conductivity of a system is the minimum value of its innate thermal conductivity. The phonons become entirely unpaired at high temperatures, and the heat energy is transferred to nearby atoms. In this case, the average interatomic distance is assumed to represent the mean free path of phonons. As a result, various atoms in a molecule can be replaced by a single atom with an average atomic mass of M/n in this approximation (n is the number of atoms in the unit cell). In a single atom, the optical mode is absent. Clarke used this concept to calculate the minimum thermal conductivity $k_{min}$ at high temperature using the formula [84]:

$$k_{min} = k_B v_m \left(\frac{nN_A\rho}{M}\right)^{2/3}$$

In this expression, $k_B$ is the Boltzmann constant, $v_m$ is the average sound velocity, $N_A$ is Avogadro's number, *r* is the mass density, *n* is the number of atoms in a molecule, and *M* is the molecular weight. The calculated values of $k_{min}$ of the $BaTiO_3$ polymorphs (Table 6) possesses a similar trend as that of the sound velocities and Debye temperature.

### 3.5.5 The Grüneisen parameter

The Grüneisen parameter γ, also known as the temperature-dependent anharmonicity factor measures the number of phonon vibrations apart from harmonic oscillation in a crystal. The lattice thermal conductivity can be confined by the phonon–phonon Umklapp and normal scattering processes, which are driven by the anharmonicity of the chemical bond [85]. The Grüneisen parameter γ of $BaTiO_3$ polymorphs has been calculated using the following relation [86]:

$$\gamma = \frac{3}{2}\left(\frac{1+\nu}{2-3\nu}\right)$$

where ν is the Poisson's ratio which can also be calculated from longitudinal ($v_l$) and transverse ($v_t$) sound waves using the relation [86,87]:

$$\nu = \frac{1-2\left(v_t/v_l\right)^2}{2-2\left(v_t/v_l\right)^2}$$

The calculated values γ of $BaTiO_3$ polymorphs are tabulated in Table 6 and their values increase in the order of cubic > hexagonal > rhombohedral > orthorhombic > tetragonal phases of $BaTiO_3$. Hence, the cubic phase displays maximum lattice thermal conductivity whereas the tetragonal phase demonstrates minimum.

### 3.5.6 Lattice thermal conductivity

The lattice thermal conductivity of a material is an essential parameter to evaluate the thermal performance at higher temperatures. When there are no free electrons to convey heat, the only route of heat transport accessible is lattice thermal conductivity. Slack devised an equation for determining the lattice thermal conductivity based on the average atomic weight of the atoms in a "molecule" (or the atoms in the formula unit of the crystal) [88]. Using Slack's model, the lattice thermal conductivity of a crystal can be calculated by using the empirical formula

$$k_{ph} = A\frac{M_{av}\Theta_D^3\delta}{\gamma^2 n^{2/3} T}$$

where $M_{av}$ is the average atomic mass of a crystal, $\Theta_D$ is the Debye temperature, $\delta$ is the cubic root of the average atomic volume, g is the Grüneisen parameter, $n$ is the number of atoms in the unit cell, and $T$ is the temperature.

The factor $A$(g) is determined by Julian's formula [89],

$$A(\gamma) = \frac{5.72\times10^7\times0.849}{2\times\left(1-\frac{0.514}{\gamma}+\frac{0.228}{\gamma^2}\right)}$$

The lattice thermal conductivity calculated for the five polymorphs of $BaTiO_3$ at room temperature (300 K) is listed in Table 6 and the temperature dependence is shown in Fig. 7.

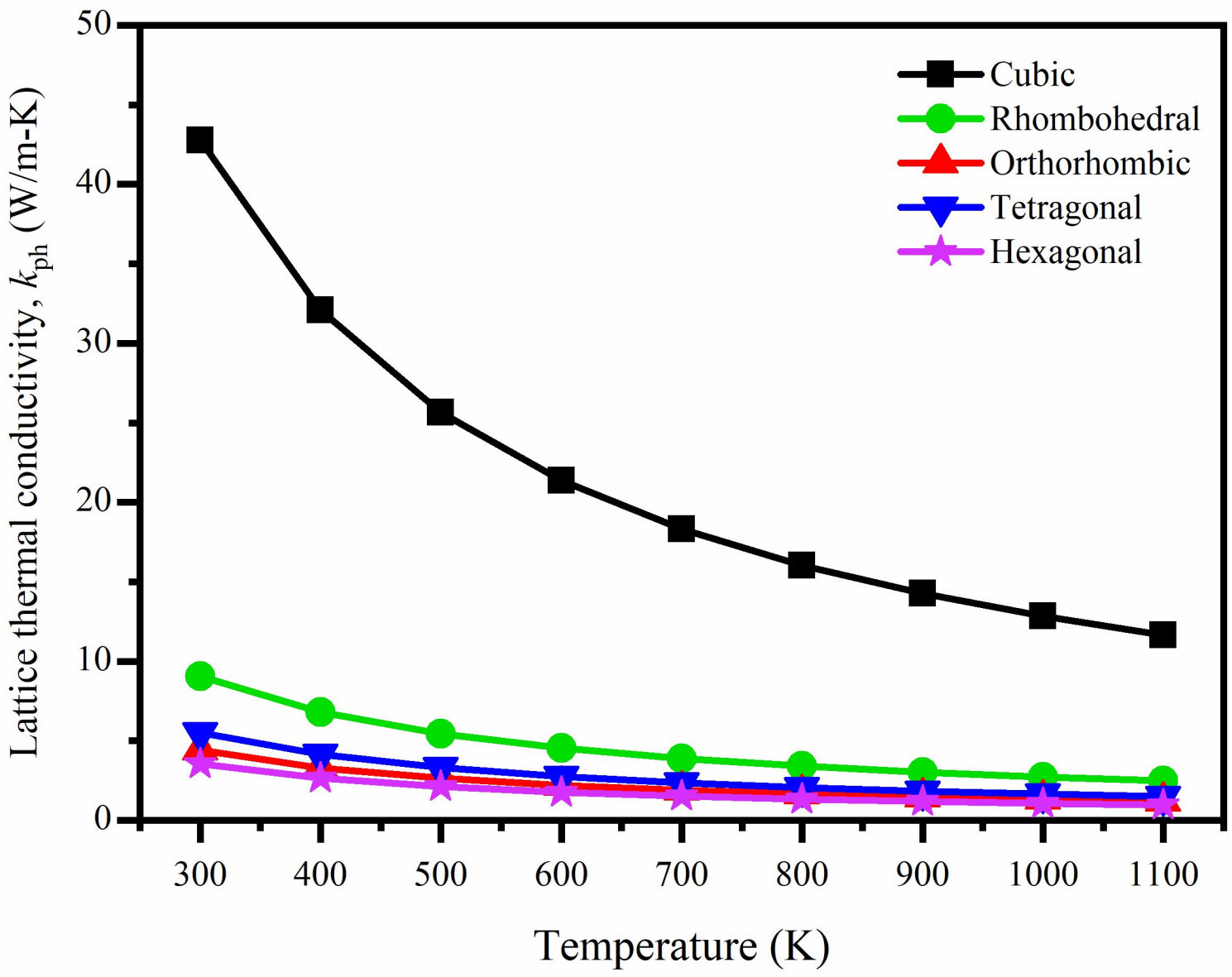

Fig. 7. Lattice thermal conductivity of the five polymorphs of $BaTiO_3$

From Fig. 7, it can be said that on the basis of the lattice thermal conductivity, the polymorphs can be separated into two groups where the cubic phase stands out from the rest and has a very high value. On the other hand, the remaining polymorphs have a small value as compared to the cubic phase. The gradual decrement of the lattice thermal conductivity with the increase of temperature is perceived for all the polymorphs. The room temperature lattice thermal conductivity follows the increasing trends of cubic > rhombohedral > tetragonal > orthorhombic > hexagonal phases of $BaTiO_3$. For thermal barrier coating application, a material $k_{min}$ should be equal to or smaller than the limiting value of 1.25 W/m-K [90], and according to this, all of the $BaTiO_3$ polymorphs should be considered as potential for TBC applications.

Table 6: The number of atoms per mole of the compound n ($m^{-3}$), minimum thermal conductivity $k_{min}$ (W/m-K), lattice thermal conductivity $k^*_{ph}$ (W/m-K) and Grüneisen parameter γ of different polymorphs of $BaTiO_3$

| Phase | n ($10^{28}$) | $k_{min}$ | $k^*_{ph}$ | γ |
|---|---|---|---|---|
| Cubic | 7.67 | 1.17 | 42.78 | 1.411 |
| Rhombohedral | 7.45 | 0.79 | 9.07 | 1.694 |
| Orthorhombic | 7.45 | 0.78 | 4.40 | 1.698 |
| Tetragonal | 7.48 | 0.79 | 5.52 | 2.147 |
| Hexagonal | 7.35 | 0.98 | 3.55 | 1.549 |

*Calculated at 300K.

### 3.6 Thermodynamic properties

Temperature and pressure dependent thermodynamic properties the $BaTiO_3$ polymorphs are calculated using the program Gibbs2 [38] and depicted in Figs. 8 (a-p). At P = 0 GPa and T = 400-500 K, phase transition occurs from the orthorhombic structure to tetragonal structure while rhombohedral to tetragonal phase transition occur at P = 0 GPa and T = 700-800 K. Therefore, only two phases exhibit phase transition at 0 GPa, on the other hand, other phases are still in same phase in the whole range of temperature at 0 GPa.

Fig. 8(a,b) shows the temperature and pressure dependence of adiabatic bulk modulus, *B* of the five polymorphs of $BaTiO_3$ at P = 0 GPa and T = 300 K, respectively. From Fig. 8(a), it is observed that B decreases slowly at a low (~100 K) temperature range, which is associated with the 3rd law of thermodynamics. After 100 K, a linear decrement was continued up to 1000 K for the cubic, tetragonal, and hexagonal phases. The bulk modulus decreases sharply in the case of the rhombohedral and orthorhombic phases. The bulk modulus, *B* increases with increasing *P* at a given *T*, which is a usual characteristic because external stress makes the compound mechanically stronger. It is also depicted that the cubic phase is the hardest at low temperatures while the hexagonal phase becomes the hardest at high temperatures. The orthorhombic phase is softer among them. The general formula for bulk modulus, $B = v\frac{\Delta p}{\Delta v}$, represents the increase of *B* with pressure *P*. From Fig. 8(b), it is evident that the bulk modulus, *B* increases linearly with increasing pressure, *P* at constant temperature *T* = 300 K for each polymorph. The orthorhombic phase has the steepest rise among them.

Fig. 8(c) presents the Debye temperature $\Theta_D$ as a function of temperature of the $BaTiO_3$ polymorphs and the value of $\Theta_D$ decreases with increasing temperature. The rate of decrease of the Debye temperature of the orthorhombic phase is higher than that of the other phases. $\Theta_D$ decreases identically for the cubic, rhombohedral, tetragonal, and hexagonal phases. The decrease of *B* and $\Theta_D$ with temperature is attributed due to the decrease of material stiffness with temperature, which leads to the decrement of *B* and $\Theta_D$ [91]. On the other hand, pressure increases the stiffness of the compound and hence increases the *B* and $\Theta_D$. Fig. 8(d) shows the calculated pressure dependence of $\Theta_D$ for the polymorphs at T = 300 K. The value of $\Theta_D$ varies in the order of orthorhombic > rhombohedral > cubic > tetragonal > hexagonal at high pressures as a function of temperature. At low temperature and pressure, both temperature and pressure dependent $\Theta_D$ of the cubic phase is higher than the other polymorphs which are analogous to the strong interatomic bonding in the crystal.

The temperature and pressure dependency of specific heat at constant volume, $C_V$, and at constant pressure, $C_P$ of the polymorphs are shown in Figs. 8(e-h). Heat capacities rise as a result of phonon thermal softening as $T$ increases. The discrepancy between $C_P$ and $C_V$ is related to thermal expansion by anharmonicity effects. $C_V$ follows the Debye equation, which states that at low temperatures (T~300 K), $C_V$ is proportional to $T^3$, and at high temperatures (T > 300 K), the anharmonic influence on heat capacity is muted and $C_V$ approaches the traditional asymptotic Dulong-Petit limit $C_V = 3nN_Ak_B = 124.72$ J/mol.K. Furthermore, $C_P$ continues to rise with temperature and for the rhombohedral and orthorhombic phases, the increases is prominent. At low $T$, $C_P$ behaves likewise to $C_V$ while $C_P$ is greater than $C_V$ at higher temperatures. From Figs. 8(g,h), it is observed that, both $C_V$ and $C_P$ decreases with increasing pressure.

Figs. 8(i,j) show the calculated results for differences in the volume thermal expansion coefficient, $\alpha_V$, as a function of temperature and pressure of the polymorphs. Except for the orthorhombic phase, $\alpha_V$ increases rapidly, especially up to T < 200 K, although it gradually tends to a moderate increase at higher temperatures. For the orthorhombic phase, the increase of $\alpha_V$ does not slow down at all. As pressure rises, the coefficient $\alpha_V$ falls exponentially. The higher value of $\alpha_V$ for the orthorhombic phase indicates that it is more malleable.

In a solid, internal energy is the result of molecules moving randomly and disorderedly. Figs. 8(k,l) display the predicted internal energy of the five phases as a function of temperature and pressure, respectively. $U$ grows slowly with temperature for $P = 0$ GPa, but at T > 100 K, it increases more quickly for all polymorphs. Furthermore, from Fig. 8(l), it is observed that the internal energy of the polymorphs increases linearly with increasing pressure at T = 300 K. It is also observed that the rate of increase of internal energy, $U$ for the hexagonal phase is slightly less than that of the other phases, in the studied pressure range.

Entropy, $S$ is a unit of measurement for how much a process causes the energy of atoms and molecules to disperse. The entropy variations of the polymorphs as a function of temperature and pressure are depicted in Figs. 8(m,n). It has been found that for the phases under consideration, the rate of entropy rise with temperature is almost identical. The orthorhombic phase has the highest entropy at $T = 300$ K and therefore has the most disordered structure. It is also observed that the entropy, $S$ decreases with increasing pressure at $T = 300$

K. This is due to the fact that, the volume decreases with increasing pressure. The rate of decrease of entropy, *S* with pressure is larger for the orthorhombic phase.

The Grüneisen parameter, γ, measures the number of phonon vibrations. Figs. 8(o,p) depict the temperature and pressure Grüneisen parameter. Fig. 8(o) depicts the effect of temperature on γ is minimal for the cubic, tetragonal and hexagonal phases whereas γ rises handsomely for the orthorhombic phase. γ decreases with pressure for all the phases as can be seen from Fig. 8(p).

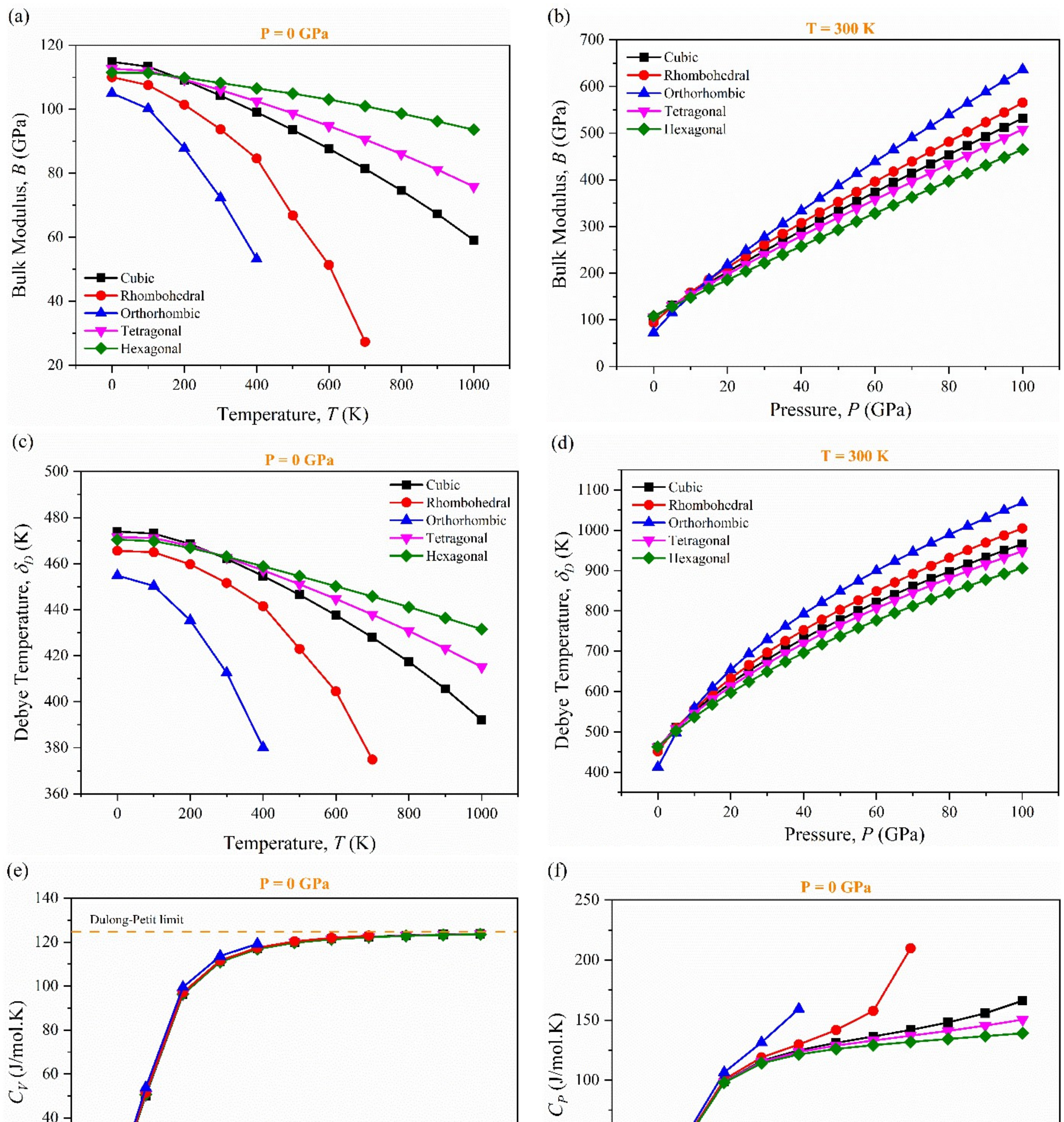

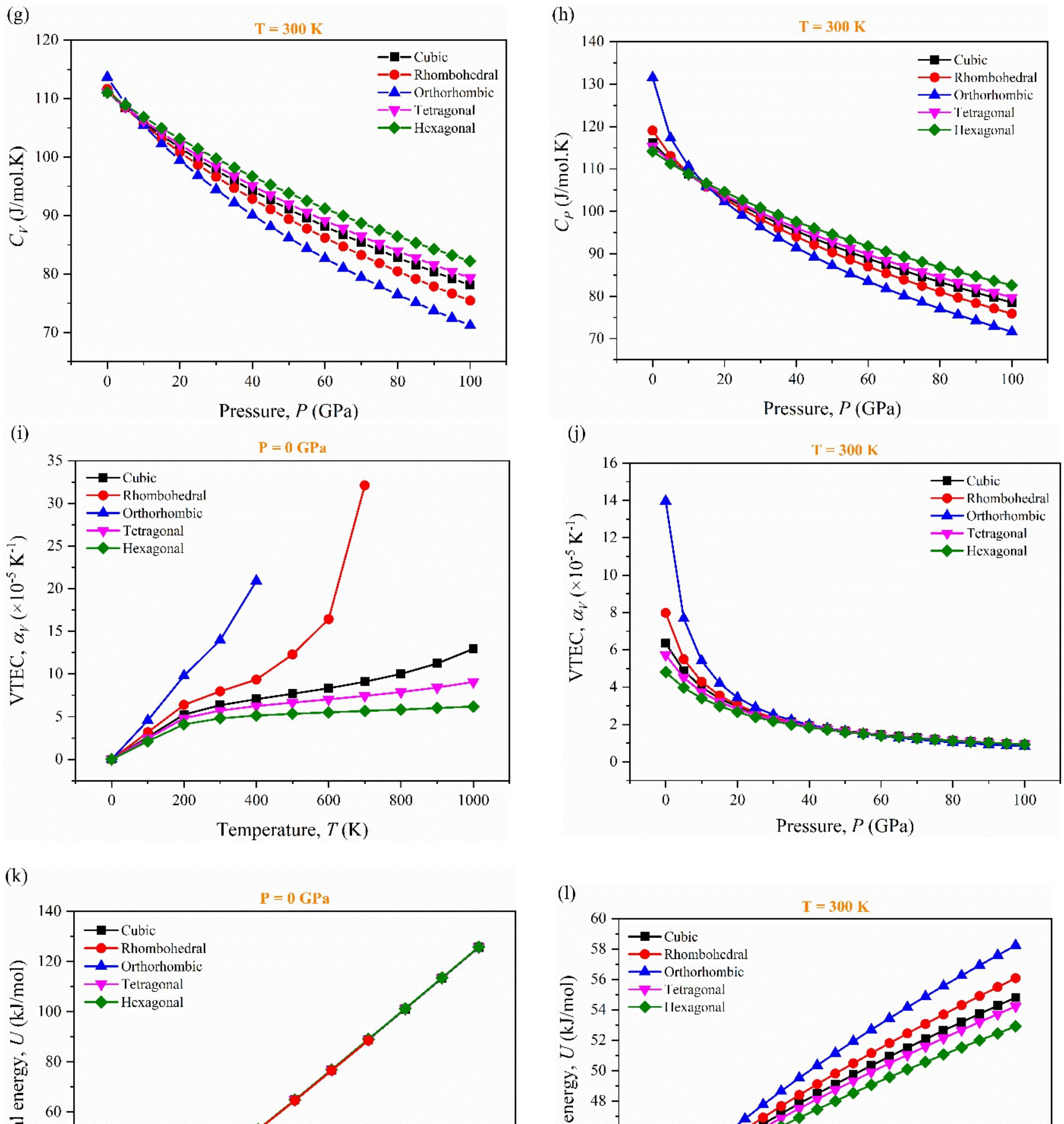

(g)
T = 300 K
Cubic
Rhombohedral
Orthorhombic
Tetragonal
Hexagonal
$C_V$ (J/mol.K)
Pressure, $P$ (GPa)
(h)
T = 300 K
Cubic
Rhombohedral
Orthorhombic
Tetragonal
Hexagonal
$C_P$ (J/mol.K)
Pressure, $P$ (GPa)
(i)
P = 0 GPa
Cubic
Rhombohedral
Orthorhombic
Tetragonal
Hexagonal
VTEC, $\alpha_V$ ($\times 10^{-5}$ K$^{-1}$)
Temperature, $T$ (K)
(j)
T = 300 K
Cubic
Rhombohedral
Orthorhombic
Tetragonal
Hexagonal
VTEC, $\alpha_V$ ($\times 10^{-5}$ K$^{-1}$)
Pressure, $P$ (GPa)
(k)
P = 0 GPa
Cubic
Rhombohedral
Orthorhombic
Tetragonal
Hexagonal
(l)
T = 300 K
Cubic
Rhombohedral
Orthorhombic
Tetragonal
Hexagonal

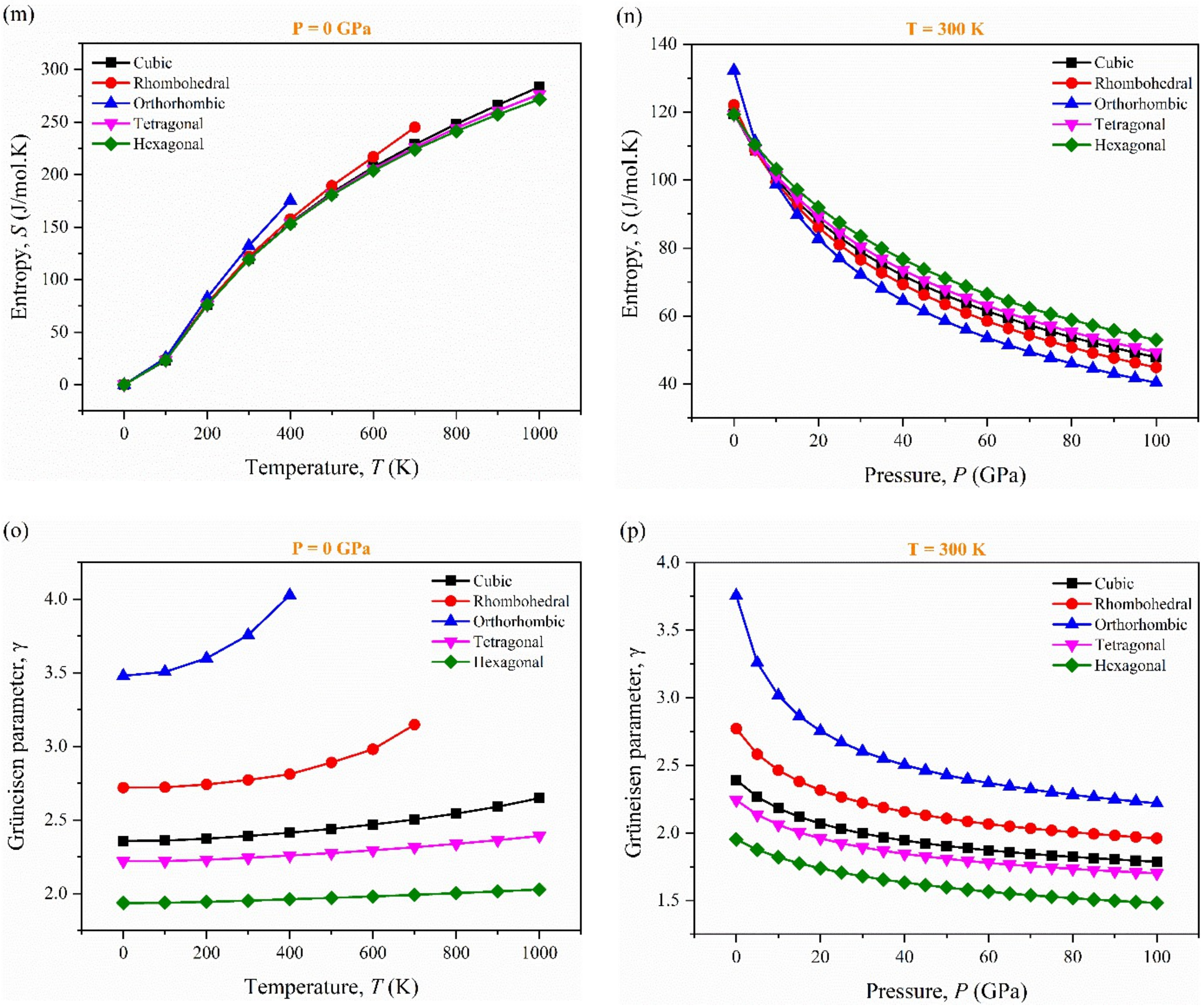
(m)
P = 0 GPa
Cubic
Rhombohedral
Orthorhombic
Tetragonal
Hexagonal
Entropy, S (J/mol.K)
Temperature, T (K)
(n)
T = 300 K
Entropy, S (J/mol.K)
Pressure, P (GPa)
(o)
P = 0 GPa
Grüneisen parameter, γ
Temperature, T (K)
(p)
T = 300 K
Grüneisen parameter, γ
Pressure, P (GPa)

Fig. 8. Temperature and pressure dependence of (a-p) bulk modulus; Debye temperature, $\Theta_D$, $\alpha_V$; specific heat at constant pressure ($C_P$) and volume ($C_V$); volume expansion coefficient, $\alpha_V$; internal energy U; entropy S; and Grüneisen parameter $\gamma$ of five polymorphs of $BaTiO_3$.

## 4. Conclusion

A wide range of mechanical and thermodynamic characteristics of $BaTiO_3$ polymorphs were analyzed, which was previously not studied in detail by employing first principles computations with GGA. The elastic constants confirm that all the polymorphs possess mechanical stability. $BaTiO_3$ polymorphs are found to be hard, machinable, and exhibit strong anisotropy with various degrees. The cubic phase is brittle while the other phases are ductile. The intermediate value of Debye temperature indicates moderate phonon thermal conductivity. It is evident from the estimated values of elastic moduli, Debye temperature, melting temperature, and thermal expansion coefficient that $BaTiO_3$ has great potential as a thermal barrier coating material (TBC). The high melting point of the polymorphs studied here demonstrates their potential applicability in thermally harsh environments. Also, the high thermal conductivity of all phases makes them suitable materials for application in the fields of thermoelectronic devices, thermal sensors, thermal exchangers, cooling systems, automobiles, and space crafts. The phonon dispersion curves and phonon density of states reveal that there is distortion. Phase transition occurs for the rhombohedral and orthorhombic phases at high temperature. The variation of $\Theta_D$ with temperature indicates that there is a strong bonding in the cubic phase as compared to the other phases. The mechanical, and thermal characteristics of the compound make it particularly well suited for engineering and device applications.

## CRediT authorship contribution statement

**Arpon Chakraborty**: Data curation, Formal analysis, Investigation, Methodology, Visualization, Writing–original draft, Writing–review & editing. **M. N. H. Liton**: Methodology, Formal analysis, Visualization, Writing–review & editing, **M. S. I. Sarker**: Writing – review & editing, **M. M. Rahman**: Writing–review & editing **M. K. R. Khan**: Conceptualization, Resources, Supervision, Writing–review & editing.

## Acknowledgments

We acknowledge the Department of Physics, University of Rajshahi, Bangladesh for providing computing facilities. Also, we gratefully acknowledge the Ministry of Science and Technology, Govt. of the People's Republic of Bangladesh for awarding the National Science and Technology (NST) fellowship.

## Declaration of competing interest

The authors declare that they have no known competing personal relationships that could have appeared to influence the work reported in this paper.

## Data availability

Data will be made available on request.

## Funding Declaration

The authors declare that they have no financial support associated with the work reported in this paper.